\documentclass[preprintnumbers,prd,twocolumn,showpacs,floatfix,nofootinbib,
superscriptaddress]{revtex4}
\usepackage[utf8]{inputenc}

\usepackage{graphicx}
\usepackage{epsfig}
\usepackage{bm}
\usepackage{amssymb}
\usepackage{float}
\usepackage{amsmath}
\usepackage{dcolumn}
\usepackage{cancel}
\usepackage[colorlinks]{hyperref}
\usepackage[usenames,dvipsnames]{color}

\begin{document}

\title{New rotating black holes   in non-linear Maxwell $f({\cal R})$ gravity}

 \author{G.G.L. Nashed}
\email{nashed@bue.edu.eg}
\affiliation {Centre for Theoretical Physics, The British University in Egypt, P.O. Box
43, El Sherouk City, Cairo 11837, Egypt}

\author{Emmanuel N. Saridakis}
\email{msaridak@noa.gr}
\affiliation{National Observatory of Athens, Lofos Nymfon, 11852 Athens,
Greece}
\affiliation{CAS Key Laboratory for Researches in Galaxies and Cosmology,
Department of Astronomy, University of Science and Technology of China, Hefei,
Anhui 230026, P.R. China}
\affiliation{School of Astronomy, School of Physical Sciences,
University of Science and Technology of China, Hefei 230026, P.R. China}

\begin{abstract}
We investigate   static and rotating  charged  spherically symmetric  solutions
in the framework of    $f({\cal R})$  gravity, allowing additionally the
electromagnetic sector to depart from linearity. Applying a convenient, dual
description for the electromagnetic Lagrangian, and using as an example the
square-root $f({\cal R})$   correction, we solve analytically the involved field
equations. The obtained solutions belong to two branches, one that contains
the   Kerr-Newman solution of general relativity as a particular limit,  and
one that arises purely from the gravitational modification with no 
general-relativity limit.
The novel black hole solution  has a true  central singularity which is hidden
behind a horizon, however for particular parameter regions the horizon
disappears and the singularity becomes a naked one.
Furthermore, we investigate the thermodynamical properties of the solutions,
such as the temperature, energy, entropy, heat capacity and Gibbs free energy.
We extract the  entropy and quasilocal energy positivity conditions, we show
that negative-temperature,  ultracold, black holes are possible, and we show
that the obtained solutions are thermodynamically stable for suitable
model parameter regions.
\end{abstract}

\maketitle

\section{ Introduction}

The detection of   gravitational waves from the binary black hole
and binary neutron star mergers by the LIGO-VIRGO collaboration
\cite{Abbott:2016blz,Abbott:2017oio,TheLIGOScientific:2017qsa} opened the new
era of multi-messenger astronomy. In this novel window to investigate the
universe the central role is played by spherically symmetric compact objects
and black holes. Since their properties are determined by the underlying
gravitational theory  recently  there has been an increased interest  in
studying such solutions in various extensions of general relativity (GR).

The simplest modification of GR arises by  generalizing the
action through arbitrary functions of the Ricci scalar, resulting to
$f({\cal R})$   gravity
\cite{DeFelice:2010aj,Nojiri:2010wj}. Nevertheless,
 one can build more complicated constructions using
  higher-order   corrections, such as the Gauss-Bonnet term $G$ and its
functions
\cite{Wheeler:1985nh,Antoniadis:1993jc,Nojiri:2005jg,
DeFelice:2008wz}, Lovelock combinations
\cite{Lovelock:1971yv,Deruelle:1989fj}, Weyl combinations
\cite{Mannheim:1988dj},
higher spatial-derivatives as in Ho\v{r}ava-Lifshitz gravity
\cite{Horava:2009uw}, etc. On the other hand, one can be based in the
teleparallel formulation of gravity, and construct its modifications such as in
$f(T)$ gravity \cite{Cai:2015emx, Bengochea:2008gz, Linder:2010py}, in
$f(T,T_G)$ gravity \cite{Kofinas:2014owa},   etc.  Hence, in all these classes
of   modified gravity one can extract the spherically symmetric black hole solutions  and study their properties
\cite{Aharony:1999ti,Awad:2005ff,Awad:1999xx,Awad:2000ac,
Cai:2012db,Anabalon:2013oea,
Cisterna:2014nua, Babichev:2015rva,Hanafy:2015yya,
Brihaye:2016lin,Cisterna:2016nwq,Babichev:2017rti,Cvetkovic:2017nkg,
Erices:2017izj,
Cisterna:2017jmv,
Cisterna:2018mww,Gonzalez:2011dr,Capozziello:2012zj,Iorio:2012cm,Nashed:2013bfa,
Aftergood:2014wla,
Paliathanasis:2014iva,2015IJMPD..2450007N,
2015EPJP..130..124N,
Junior:2015fya,Kofinas:2015hla,Nashed:2016tbj,Cruz:2017ecg,Awad:2017sau,
Ahmed:2016cuy, Farrugia:2016xcw,Koutsoumbas:2018gbd,
Mai:2017riq,Doneva:2018rou,
Awad:2017tyz,
Bejarano:2017akj,Abdalla:2019irr,Nashed:2018cth,Destounis:2019omd,
Papantonopoulos:2019ugr}.

In general, the obtained spherically symmetric solutions can be classified
either in branches which are extensions of the corresponding GR solutions,
coinciding exactly with them in a particular limit, or to novel branches that
appear purely from the gravitational modification and do not possess a GR
limit.  In both cases, the obtained black holes and compact objects present new
properties which may be potentially detectable in the gravitational waves
arising from mergers. Thus, studying the properties of    spherically symmetric
solutions in various modified gravities is crucial in order to put the new
observational tool of multi-messenger astronomy to work.

It is the aim of the present study to derive  new charged black hole
solutions in the context of   $f({\cal R})$ gravity, allowing additionally for
possible non-linearities in the Maxwell sector.
 The plan of the manuscript   is as follows:
 In Section \ref{S2},  we
present a convenient way to handle the possible electrodynamic non-linearities.
In Section \ref{S3} we extract static and rotating spherically symmetric
black hole solutions and in Section
  \ref{S66}  we calculate all the thermodynamical quantities such as the
entropy, Hawking temperature, heat capacity and Gibb's free energy, analyzing
additionally the stability of the solutions.
Finally,   Section \ref{S77} is devoted to discussion and
conclusions.

\section{Dual representation of non-linear electrodynamics }\label{S2}

 In this section we present a new way for the description of non-linear
electrodynamics, which is valid independently of the specific electromagnetic
Lagrangian and which allows for an easy handling concerning the derivation of
field equations. We start with a general gauge-invariant electromagnetic
Lagrangian of the form  ${\cal L({\cal F})}$, where ${\cal F} =
\frac{1}{4}{\cal F}_{\alpha \beta}{\cal F}^{\alpha \beta}$ is
the usual anti-symmetric Faraday
tensor  defined as $ {\cal F}_{\alpha \beta}=2 {V}_{[\alpha, \beta]},$
with ${V}_\mu$   the gauge
potential 1-form and where
  square brackets denote symmetrization
\cite{plebanski1970lectures}.
 As usual, linear, Maxwell electrodynamics is obtained for    ${\cal L({\cal
F})}=4{\cal F}$.

For the purposes of this work we consider a dual  representation, introducing
     the auxiliary field ${\cal P}_{\alpha \beta}$, which has been proven
convenient if one desires to embed electromagnetic in the framework of general
relativity  \cite{1999PhLB..464...25A,1987JMP....28.2171S}. In particular, we
impose the   Legendre transformation
\begin{equation} \label{Ha}
{\mathbb{H}}=2{\cal F} {\cal L}_{\cal F}-{\cal L},
\end{equation}
with $
{\cal L}_{\cal F}\equiv\frac{\partial {\cal L}}{\partial {\cal F}}$.
Defining
\begin{eqnarray}
\label{rel22}
{\cal P}_{\mu \nu}={\cal L}_{\cal F} {\cal
F}_{\mu \nu},
\end{eqnarray}
we immediately find that   ${\mathbb{H}}$ is an arbitrary function of the
invariant
\begin{eqnarray}
\label{rel22vvv}
{\cal P}=\frac{1}{4}{\cal P}_{\alpha \beta}{\cal
P}^{\alpha \beta}={\cal L}_{\cal F}^2 {\cal
F} .
\end{eqnarray}
Using  (\ref{Ha}), the Lagrangian of   non-linear
electrodynamics can be represented  in terms of ${\cal P}$ as
\begin{equation} \label{rel11}
{\cal L}=2{\cal P} {\mathbb{H}}_{\cal
P}-{\mathbb{H}},
\end{equation}
while
\begin{eqnarray}
\label{rel22}
{\cal F}_{\mu \nu}={\mathbb{H}}_{\cal P} {\cal
P}_{\mu \nu},  \end{eqnarray}
with ${\mathbb{H}}_{\cal
P}=\frac{\partial {\mathbb{H}}}{\partial {\cal P}}$.

 The field equations  thus acquire the form
\cite{1999PhLB..464...25A}
\begin{equation} \label{maxf}
\partial_\nu \left( \sqrt{-g} {\cal P}^{\mu \nu} \right)=0,\end{equation}
and the corresponding energy-momentum tensor   is given as
\begin{equation} \label{max1}
{{{\mathfrak{
T}}^{{}^{{}^{^{}{\!\!\!\!\scriptstyle{nlem}}}}}}}^\nu_\mu\equiv
2({\mathbb{H}}_{\cal P}{\cal P}_{\mu \alpha}{\cal P}^{\nu \alpha}-\delta_\mu^\nu
[2{\cal P}{\mathbb{H}}_{\cal P}-{\mathbb{H}}]).
\end{equation}
We mention that in   general     (\ref{max1}) has  a non-vanishing
  trace
  \begin{equation}
  {{{\mathfrak{
T}}^{{}^{{}^{^{}{\!\!\!\!\scriptstyle{nlem}}}}}}}=8({\mathbb{H}}-{\mathbb{H}}_{
\cal P}{\cal P}),
\label{traceEM}
\end{equation}
which becomes zero only in the case of the linear theory.
  Finally, the electric and magnetic fields in spherical coordinates
can be calculated as  \cite{1999PhLB..464...25A,1987JMP....28.2171S}
\begin{eqnarray} \label{Max3}
&&E = \int{\cal F}_{tr}dr =\int{\mathbb{H}}_{\cal P}{\cal P}_{tr}dr,
\nonumber\\
&& B_r =\int {\cal F}_{r\phi}d\phi =\int{\mathbb{H}}_{\cal P}{\cal
P}_{r\phi}d\phi,
\nonumber\\
&&
B_\theta = \int{\cal F}_{\theta r}dr
=\int{\mathbb{H}}_{\cal P}{\cal P}_{\theta r}dr, \nonumber\\
&& B_\phi=\int{\cal F}_{\phi r}dr =\int{\mathbb{H}}_{\cal P}{\cal P}_{\phi r}dr.
\end{eqnarray}\\

\section{Static and rotating black hole solutions in   non-linear Maxwell
$f({\cal R})$ gravity
}\label{S3}

In this section we consider non-linear electrodynamics in a gravitational
background governed by  $f({\cal R})$ gravity, and we extract charged black
hole solutions. The total action is written as
  \cite{Carroll:2003wy}
  \begin{equation} \label{act}
  S_{t}=\frac{1}{2\kappa}\int \sqrt{-g}f({\cal
R})~d^{4}x+\int \sqrt{-g}{\cal
L({\cal F})}~d^{4}x,\end{equation}
 where $\sqrt{-g}$ is the determinant of the metric $g_{\mu\nu}$ and
$\kappa$  is  the gravitational constant (from now on we set $\kappa=1$ and
all quantities are measured in these units).
Performing variation with respect to the metric leads to the gravitational
field equations
\cite{Cognola:2005de,Koivisto:2005yc}:
\begin{eqnarray} \label{f1}
&&
\!\!\!\!\!\!\!\!\!\!\!\!\!\!\!\!
\xi_{\mu \nu}={\cal R}_{\mu \nu} f_{\cal R}-\frac{1}{2}g_{\mu \nu}f({\cal
R})-2g_{\mu \nu}\Lambda +g_{\mu \nu} \Box f_{\cal R}\nonumber\\
&&
-\nabla_\mu
\nabla_\nu
f_{\cal R}-8\pi {{{\mathfrak{
T}}^{{}^{{}^{^{}{\!\!\!\!\scriptstyle{nlem}}}}}}}_{\mu\nu}\equiv0,
\end{eqnarray}
where $\Box$ is the D'Alembertian operator     defined as $\Box=
\nabla_\alpha\nabla^\alpha $, $\nabla_\alpha V^\beta$ is the covariant
derivative  of the vector $V^\beta$,  $f_{\cal R}\equiv\frac{df({\cal
R})}{d{\cal R}}$, and the electromagnetic energy-momentum tensor ${{{\mathfrak{
T}}^{{}^{{}^{^{}{\!\!\!\!\scriptstyle{nlem}}}}}}}_{\mu\nu}$ is given by
(\ref{max1}).
Additionally, taking the trace of  (\ref{f1}) gives
\begin{eqnarray} \label{f3}
\xi={\cal R}f_{\cal R}-2f({\cal R})-8\Lambda+3\Box f_{\cal R}-{{{\mathfrak{
T}}^{{}^{{}^{^{}{\!\!\!\!\scriptstyle{nlem}}}}}}},
\end{eqnarray}
with $
{{{\mathfrak{
T}}^{{}^{{}^{^{}{\!\!\!\!\scriptstyle{nlem}}}}}}}$ given by (\ref{traceEM}).

\subsection{Static solutions}

In order to extract   black hole solutions  we consider
a spherically symmetric metric of the form
\begin{eqnarray} \label{met}
  ds^2=H(r)dt^2-\frac{dr^2}{H(r)}-r^2(d\theta^2+\sin^2\theta d\phi^2).
\end{eqnarray}
Thus, the corresponding
  Ricci scalar  becomes
\begin{eqnarray}\label{r1}
 { \cal R}=\frac{2-r^2H''-4rH'-2H}{r^2},
\end{eqnarray}
where from now on primes denote derivatives with respect to $r$.
Concerning the electromagnetic
potential 1-form we consider the general ansatz \cite{Elizalde:2020icc}
\begin{equation}
\label{pot1} V :=q(r)dt+n(\phi)dr+s(r)d\phi,
\end{equation}
with  $q(r)$,$s(r)$,$n(\phi)$   three free functions
reproducing
 the electric and magnetic charges in the vector potential where $ {\cal P}=dV$ and $V
= V_\nu dx^\nu$.

In the following, without loss of generality, and just to provide an example
of the method at hand, we focus on the square-root
$f({\cal R})$ correction to general relativity, where   $f({\cal R})={\cal
R}-2\alpha\sqrt{{\cal R}}$
\cite{Dimitrijevic:2019pct,Nashed:2019tuk}.
 Inserting the  metric (\ref{met}) and the potential
(\ref{pot1})
into the general
field equations (\ref{f1}), (\ref{f3}), (\ref{maxf}) we obtain the following
non-vanishing field equations:
\begin{widetext}
\begin{eqnarray} \label{df1}
& &
\!\!\!\!\!\!\!\!\!\!\!\!\!\!\!\!
\xi_t{}^t=\frac{1}{4r^6\sqrt{R^5}} \Big\{2R r^6
HH''''+3r^6HH'''^2+r^3H'''[r^2H''(12H-rH')-4r^2H'^2-2r(31H-1)H'+48H(1-H)]
\nonumber\\
& &
\ \ \ \ \ \ \ \ \ \ \ \
+2r^6H''^3
+4r^4H''^2(6rH'+15H-4)+2r^2H''[57r^2H'^2+14rH'(57H-5)+4(5+4H-3H^2)]
\nonumber\\
& &\ \ \ \ \ \ \ \ \ \ \ \
+200r^3H'^3+4r^2H'^2(96H-85)
+8rH'(H-1)(27H-23)+32(H-1)^2(2H-1)\Big\}
\nonumber\\
& &
-\frac{2rsin^2\theta q'q''(2r^2\mathbb{H}-rH'+1-H)}{2r^3\sin^2\theta q'q''
+(s'-n_\phi)[2Hs''+(s'-n_\phi)(rH'-2H)]}\nonumber\\
& &
-\frac{(s'\!-\!n_\phi)\{2rHs''(1\!-\!rH'\!-\!H-\mathbb{H}r^2)-(s'\!-\!n_\phi)[
r^2H'^2\!-\!r(1\!+\!H+2r^2\mathbb{H})]+2H(H\!-\!1\!-\!r^3\mathbb{H}
'\!-\!r^2\mathbb{ H} )\} }{ 2r^2[ 2rHs''(s'-n_\phi)+2r^3\sin^2\theta q'q''
+(s'-n_\phi)^2(rH'-2H)]}=0,
\end{eqnarray}
\begin{eqnarray}\label{df2}
& &
\!\!\!\!\!\!\!\!\!\!\!\!\!\!\!\!
 \xi_t{}^\phi=\frac{4rH\mathbb{H}'q'(s'-n_\phi)}{ 2r^3\sin^2\theta q'q''
+(s'-n_\phi)[2Hs''+(s'-n_\phi)(rH'-2H)]}=0, \ \ \ \ \  \ \ \ \ \  \ \ \ \ \  \ \
\ \ \  \ \ \ \ \  \ \ \ \ \  \ \ \ \ \  \ \ \ \ \  \ \ \ \ \  \ \ \ \ \  \ \ \ \
\  \ \ \ \ \
\end{eqnarray}
\begin{eqnarray}\label{df3}
& &
\!\!\!\!\!\!\!\!\!\!\!
\!\!\!\!\!\!\!\!\!\!\!\!\!
\xi_r{}^r=\frac{1}
{4r^4\sqrt{R^3}}\Big\{4r^2\sqrt{R^3}(2r^2\mathbb{H}-rH'-H+1)+\alpha\Big[
r^3H'''(rH'+4H)-2r^4H''^2+4r^2H''[(3+H)-4rH']\nonumber\\
& &
\ \ \ \ \  \ \ \ \ \  \ \ \ \ \   \ \ \ \ \  \ \ \ \ \  \ \ \ \ \  \ \ \ \ \   \
\ \ \ \  \ \ \ \ \  \ \ \ \ \  \ \ \ \ \   \
-50r^2H'^2
+4rH'(15-17H)-16(1+2H^2-3H)\Big]\Big\}=0,
\end{eqnarray}
\begin{eqnarray}\label{df4}
& &
\!\!\!\!\!\!\!
\!\!\!\!\!\!\!\!
\xi_\theta{}^\theta=\frac{1}{4r^6\sqrt{R^5}\{2r^3\sin^2\theta q'q''
-(s'-n_\phi)[2Hs''+(s'-n_\phi)(rH'-2H)]\}}\nonumber\\
&&
\cdot
\Big\{r^5\sqrt{R^5}
\Big\{rH''\{2r^3\sin^2\theta q'q''
-(s'-n_\phi)[2Hs''+(s'-n_\phi)
(rH'-2H)]\}-\frac{1}{2}rH'^2(s'-n_\phi)^2-2r^4\mathbb{H}'q'^2\sin^2\theta
\nonumber\\
&
&
\ \ \ \ \  \ \ \ \ \ \ \ \
-[q'q''r^3\sin^2\theta-2rHs''(s'-n_\phi)](2r\mathbb{H}
-H')-(H+r^2\mathbb{H})[
(s'-n_\phi)^2-2rH(\phi'^2-2s'n_\phi+s'^2)]
\Big\}
\nonumber\\
&& \ \
-\alpha\{2r^3\sin^2\theta q'q''
-(s'-n_\phi)[2Hs''+(s'-n_\phi)(rH'-2H)]\}\nonumber\\
&&\ \,
 \cdot
\Big\{2r^6RHH''''-3r^6HH'''^2+2r^3H'''[
r^2H''(rH'-7H)+4r^2H'^2+2rH'(14H-1)-22H(1-H)]\nonumber\\
& & \ \  \ \ \,
-4r^4[r^2H''^3+H''^2(18H+9rH'-5)]
-4r^2H''[33r^2H'^2+rH'(27H-34)+10H+8-18H^2]\nonumber\\
&
&
\ \  \ \ \,
 -208r^3H'^3-4r^2H'^2(81H-74)+8r(15H-16)(1-H)+16(1-4H+5H^2-2H^3)\Big\}
\Big\}\equiv0,
\end{eqnarray}
\begin{eqnarray}\label{df5}
&&
\!\!\!\!\!\!\!\!
\!\!\!\!\!\!\!\
\xi_\phi{}^\phi=\frac{1}{4r^6\sqrt{R^5}\{2r^3\sin^2\theta
q'q'' -(s'-n_\phi)[2Hs''+(s'-n_\phi)(rH'-2H)]\}}\nonumber\\
&&  \ \
\cdot
\Big\{2r^5\sqrt{R^5}  \Big\{rH''\{2r^3\sin^2\theta q'q''
-(s'-n_\phi)[2Hs''+(s'-n_\phi)
(rH'-2H)]\}
- [  rH'^2/2-(H+r^2\mathbb{H})](s'-n_\phi)^2
\nonumber\\
&& \ \ \ \ \  \ \ \ \ \ \ \ \  \ \ \ \
-r[q'q''r^2\sin^2\theta+2Hs''(s'-n_\phi)](2r\mathbb{H}
-H')-2r[H\mathbb{H}
(\phi'^2-2s'n_\phi+s'^2)
-r^3\mathbb{H}'q'^2\sin^2\theta]\Big\}\nonumber\\
&& \ \ \ \ \
+\alpha\{2r^3\sin^2\theta
q'q''
-(s'-n_\phi)[2Hs''+(s'-n_\phi)(rH'-2H)]\}
\nonumber\\
&&
\ \ \ \ \
\,\cdot
\Big\{2r^6RHH''''+3r^6HH'''^2
-2r^3H'''[r^2H''(rH'-7H)+4r^2H'^2+2rH'(14H-1)-22H(1-H)]\nonumber\\
& &\ \ \ \ \ \ \ \
+4 r^4[r^2H''^3+H''^2(18H+9rH'-5)]
+4r^2H''[33r^2H'^2+rH'(27H-34)+10H+8-18H^2]
\nonumber\\
&
&\ \ \ \ \ \ \ \
+208r^3H'^3+4r^2H'^2(81H-74)-
8rH'(15H-16)(1-H)-16(
1-4H+5H^2-2H^3)\Big\}
\Big\}\equiv0,
\end{eqnarray}
\begin{eqnarray}\label{df6}
& &
\!\!\!\!\!\!\!\!\!\!\!\!\!\!\!\!\!\!\!\!\!\!\!\!\!\!\!\!\!\!\!\!\!\!\!\!\!\!\!\!
\!\!\!\!\!\!\!\!\!\!\!\!\!\!
\!\!\!\!\!\!\!\!\!\!\!\!\!\!\!\!
\xi_\phi{}^t=\frac{4r^3\mathbb{H}'\sin^2\theta
q'(s'-n_\phi)}{2rHs''(s'-n_\phi)-2r^3\sin^2\theta q'q''
+(s'-n_\phi)^2(rH'-2H)}.\ \ \ \ \ \ \ \ \ \ \ \ \ \ \ \ \ \ \ \ \ \ \ \ \ \ \ \
\ \ \ \ \ \ \ \
\end{eqnarray}
\end{widetext}
Finally, the trace equation (\ref{f3}) becomes
\begin{widetext}
\begin{eqnarray}
& &
\!\!\!\!\!\!\!\!\!\!\!\!\!\!\!
\xi=\frac{1}{2r^6\sqrt{R^5}\{2r^3\sin^2\theta q'q''
-(s'-n_\phi)[2Hs''+(s'-n_\phi)(rH'-2H)]\}}\nonumber\\
&&\!\!\!\!\!
\cdot
\Big\{r^4\sqrt{R^5}
\Big\{r^2H''\{2r^3\sin^2\theta q'q'' -(s'-n_\phi)[2rHs''+(s'-n_\phi)
(rH'-2H)]\} +4r^5\mathbb{H}
'q'^2\sin^2\theta \nonumber\\
&&
\ \ \ \ \ \ \ \ \ \ \
-r(1+4r^2\mathbb{H}-2rH'-H)[
2q'q''r^2\sin^2\theta-2Hs''(s'-n_\phi)]-[2r^2H'^2-r(3H+4r^2\mathbb{H}+1)]
(s'-n_\phi)^2\nonumber\\
&&
\ \ \ \ \ \ \ \ \ \ \
-2H[\phi'^2-2s'n_\phi+s'^2](1+2r^3\mathbb{H}'-H+4r^2H)\Big\}\nonumber\\
&&
+\alpha\{2r^3\sin^2\theta q'q''
-(s'-n_\phi)[2Hs''+(s'-n_\phi)(rH'-2H)]\}
\nonumber\\
& &
\cdot
\Big\{2r^6RHH''''+3r^6HH'''^2-2r^3H'''[
r^2H''(rH'-6H)+4r^2H'^2+2rH'(16H-1)
-24H(1-H)]\nonumber\\
&&\ \, \
+4r^6H''^3+2r^4H''^2(3-17H-5rH')-4r^2H''[41r^2H'^2-2rH'(16H'-23)+4(
H+3-4H^2)
]\nonumber\\
&&\ \ \,
-136r^3H'^3
+2r^2H'^2(106-117H)+8rH'(15H-13)(1-H)
 -16(2H-1)(1-H)^2\Big\}\Big\}\equiv0.\label{dftrace}
\end{eqnarray}
\end{widetext}

  From Eqs.  (\ref{df2})  and (\ref{df6}) we acquire
 \begin{eqnarray} \label{sols}
 s(r)=c_4 r,\quad n(\phi)=c_4 \phi.
 \end{eqnarray} 
  Inserting   (\ref{sols}) into Eqs. (\ref{df4}) and (\ref{df5}) it is 
easy to show that two of the other
equations coincide, namely $\xi_\theta{}^\theta=\xi_\phi{}^\phi$. Therefore,
the system of differential equations  (\ref{df1}), (\ref{df3}), (\ref{df4}) and 
(\ref{df5})reduces to three
differential equations of three unknowns, $H(r)$, $\mathbb{H}$ and $q(r)$, 
which can be  solved to give the following analytical solutions: 
\begin{eqnarray} \label{sol11}
& &
\!\!\! \! \! \!
 -H(r)=\frac{c}{2}+\frac{c_1}{r}+\frac{c_2}{r^2}\,,  \qquad \qquad 
 q(r)=\frac{c_3}{r}\,, \nonumber\\
&&\!\!\! \!\!\!
 \mathbb{H}(r)=\frac{\alpha[3 c_2+4(c-1)r]r^2+\sqrt{2-c}[(c-2)r^2-2c_1]}{4
\sqrt{2-c}r^4}\,,  \nonumber\\ \label{sol11b}
 \end{eqnarray}
\begin{eqnarray}
\label{sol22}
&&
\!\!\! \! \! \!   - H(r)=\frac{c}{2}-\frac{1}{3\alpha
r}-\frac{1}{3\alpha r^2}\,,  \qquad \qquad  q(r)=\frac{c_3}{r}\,, \nonumber\\
&&\!\!\! \!\!\!   \mathbb{H}(r)=\frac{\alpha[-3+12(c-1)r]
r^2+\sqrt{2-c}[3\alpha (c-2)r^2+2]}{12 \sqrt{2-c}r^4}\,.  \nonumber\\
\label{sol22b}
\end{eqnarray}
 We stress that we adjust the constants $c_1$
and $c_2$ so that solutions  (\ref{sol11}) and (\ref{sol22}) satisfy the trace 
Eq. (\ref{dftrace}) too, and hence the whole solution structure is consistent. 
Additionally, concerning the parameter $c$ we deduce that it must be 
non-negative in order to maintain the metric signature and also the value   0 is 
excluded too in order to obtain
asymptotic flat spacetime at $r\rightarrow\infty$. If   $\alpha=0$  
then solution (\ref{sol11}) holds for any $0<c$ while solution (\ref{sol22}) 
does not exist, nevertheless if $\alpha\neq0$ then we should restrict $c$ to  
$0<c<2$ in both solutions in order to acquire real $ {\mathbb{H}(r)}$.

Concerning the function ${\cal P}$ we find ${\cal P}(r)=\frac{c_3^2}{2 r^4}$.
Hence, knowing $\mathbb{H}(r)$   we can find that
\begin{eqnarray} \label{eqHP1}
& &
\!\!\!\!\!\!\!\!\!\!\!\!\!\!\!\!\!\!
\mathbb{H}({\cal P})=
\frac{{\cal P}}{4 c_3^2}\left
\{
-4c_1+\sqrt{2}(c-2)c_3 {\cal P}^{-1/2}\right. \nonumber\\
&&
\!\!\!\!\!\!\!\!\!\!\!\!\!\!
\left.
+\alpha
c_3[{\cal
P}(1\!-\!c/2)]^{-1/2}\!\left[3c_2\!+\!2^{7/4}(c\!-\!1)\sqrt{c_3} {\cal
P}^{-1/4}\right ]\!
\right\}.
\end{eqnarray}
Therefore, knowing from (\ref{Ha}) that ${\mathbb{H}}=2{\cal F} {\cal L}_{\cal
F}-{\cal L}$ and from
(\ref{rel22vvv}) that ${\cal P} ={\cal L}_{\cal F}^2
{\cal F}$, we can re-write (\ref{eqHP1}) as
\begin{eqnarray} \label{eqHP1b}
&&
\!\!\!\!\!\!\!\!\!\!\!\!\!\!\!\!\!\!\!\!
2{\cal F} {\cal L}_{\cal
F}-{\cal L}=
\frac{{\cal L}_{\cal F}^2
{\cal F}}{4 c_3^2}\left
\{
-4c_1+\sqrt{2}(c-2)c_3  {\cal L}_{\cal F}^{-1} {\cal F}^{-1/2}\right.
\nonumber\\
&&\ \ \ \ \ \
\left.
+\alpha
c_3{\cal L}_{\cal F}^{-1}[{\cal
F}(1-c/2)]^{-1/2}\right.
\nonumber\\
&&
\ \ \ \ \ \
\left.
\cdot\left[3c_2+2^{7/4}(c-1)\sqrt{c_3} {\cal L}_{\cal F}^{-1/2}
{\cal F}^{-1/4}\right ]
\right\},
\end{eqnarray}
which is a differential equation for ${\cal L}({\cal F})$.

According to the value of $c$ the solution of the above differential equation 
will give a  corresponding correction to the standard electromagnetic 
Lagrangian. Since for our novel solution we have $0<c<2$, in the rest of the 
work we focus on the case    $c=1$, since this leads to the simple solution (but 
still a novel
solution comparing to general relativity)
\begin{equation} \label{LFsol1}
 {\cal L}({\cal F})=\frac{c_3^2}{c_1} {\cal F}+ \frac{c_3 }{\sqrt{c_1}}
\sqrt{{\cal F}} c_4 + c_1 c_4^2 .
\end{equation}
  As one can see, the first term is
standard linear electromagnetism, while the second term is the non-linear
correction of a square-root form. In the general $c$ case
the correction terms  take more complicated forms.

Let us analyze the properties of the obtained spherically symmetric solutions
(\ref{sol11}), (\ref{sol22}). These can be re-written in the standard form as
\begin{eqnarray} \label{me}
&&
ds_1{}^2=-\left(\frac{c}{2}-\frac{2M}{r}+\frac{q^2}{r^2}
\right)dt^2\nonumber\\
&&
\ \ \ \ \ \ \   \  \ \,
+\frac{dr^2}{\left(\frac{c}{2}-\frac{2M}{r}+\frac{q^2}{r^2
}\right)}+r^2d\Omega^2\;, \nonumber\\
&&\textrm{where} \qquad M=-\frac{c_1}{2}, \qquad
q=\sqrt{c_2},
\end{eqnarray}
for (\ref{sol11}), and
\begin{eqnarray}
\label{me2}
&&ds_2{}^2=-\left(\frac{c}{2}-\frac{2M}{r}+\frac{q^2}{r^2}\right)dt^2\nonumber\\
&&
\ \ \ \ \ \ \   \  \ \,
+\frac{dr^2
}{\left(\frac{c}{2}-\frac{2M}{r}+\frac{q^2}{r^2}\right)}+r^2d\Omega^2\;,
\nonumber\\
&&\textrm{where} \qquad M=\frac{1}{6\alpha}, \qquad
q=\frac{1}{\sqrt{6\alpha}},
\end{eqnarray}
for (\ref{sol22}).
The first solution branch includes the GR solution
 in the limit $\alpha\rightarrow0$ and $c\rightarrow2$ (solution
(\ref{sol11})  is
the generalization  of those  obtained in
\cite{Sebastiani:2010kv} in the static and non-charged case, see also
\cite{Nashed:2013bfa,2012ChPhL..29e0402G,Nashed:2014sea,Nashed:uja}). On the
other hand,  the
second branch    exists only in the case $\alpha\neq0$, and thus it is a novel
solution that arises
from the $f({\cal R})$ gravitational modification as well as from the
electrodynamic non--linearity.  
Hence, the two solutions, although looking similar, they are fundamentally
different, and the fact that
the mass of solution (\ref{me2}) depends only on $ 1/\alpha$ is a reflection of
the novelty of the solution (such a connection between the gravitational
modification parameters with the black hole  quantities, in specific exact 
solutions, is known to be the case in many modified gravity theories).   In 
this work  we are
interesting in solution (\ref{me2}), i.e (\ref{sol22}),  exactly because
it is a novel one with no general relativity limit. 

In order to investigate the horizons and singularities of the above solutions
   we  calculate   the  Kretschmann, the Ricci tensor  square and  the
Ricci invariants. For (\ref{sol11}) we find
\begin{eqnarray} \label{scal1}
&&
\!\!\!\!\!\!
{\cal R}^{\mu \nu \lambda \rho}{\cal R}_{\mu \nu \lambda \rho}=
r^{-8}\Big\{8c_1(7c_1-r^2)+4c_1r(12c_2+c r)
\nonumber\\
&&\ \ \ \ \
+c
r^4(c-4)+4c_2r^2(3c_2-2r)+4r^3(r+c_2c)\Big\},
\nonumber\\
&&
\!\!\!\!\!\!
{\cal R}^{\mu \nu}{\cal R}_{\mu \nu}=\frac{8c_1(c_1+r^2)+c r^2(c
r^2-4c_1)+4r^4(1-c)}{18\alpha^2r^8},
\nonumber\\
&&\!\!\!\!\!\!
{\cal R}=
\frac{2-c}{r^2},\nonumber\\
\end{eqnarray}
while for (\ref{sol22}) we acquire
\begin{eqnarray}
 && \!\!\!\!\!\! \!\!\!
 -{\cal R}^{\mu \nu \lambda \rho}{\cal R}_{\mu \nu \lambda
\rho}=
 (9\alpha^2r^8)^{-1}
\Big\{56+9r^4 \alpha^2[c-2]^2\nonumber\\
&&\ \ \ \ \ \ \ \ \ \
-12\alpha
r^3(c-2)-12r^2[\alpha(c-2)-1]+48r\Big\}, \nonumber\\
 && \!\!\!\!\!\! \!\!\!
{\cal R}^{\mu \nu}{\cal R}_{\mu
\nu}=\frac{9r^4\alpha^2(c-2)^2+12\alpha
r^2(c-2)+8}{18\alpha^2r^8}, \nonumber\\
 && \!\!\!\!\!\! \!\!\!
{\cal R}= \frac{2-c}{r^2}.
 \label{scal2}
\end{eqnarray}
 Expressions (\ref{scal1}), (\ref{scal2}) reveal that the spherically symmetric
solutions
 exhibit  a true singularity at $r=0$. Although in the GR case this
singularity is always hidden by a horizon,  when the $f({\cal R})$
 correction is switched on  this is not always the case, namely a naked
singularity may appear. This issue will be investigated in the next section.

 Lastly, concerning the electric and magnetic
  charges, expressions (\ref{Max3}) give
 \begin{eqnarray} \label{me1}
 &&
 \!\!\!\!\!\! \!
 E=\frac{[(c\! -\! 2)r^2-8m]\sqrt{2\! -\! c}+r^2\alpha
[(c\! -\! 1)r\! +\! 3q^2]}{4c_3r\sqrt{2\! -\! c}},\nonumber\\
&& \!\!\!\!\!\! \!
B_\phi=\frac{c_4r([2(c\! -\! 2)r^2+16m]\sqrt{2\! -\!
c}+3r^2\alpha
[(c\! -\! 1)\! +\! 2q^2])}{24c_3{}^2\sqrt{2\! -\! c}}, \nonumber\\
&& \!\!\!\!\!\! \!
B_r=\frac{c_4\phi([(c\! -\! 2)r^2+8m]\sqrt{2\! -\! c}+r^2\alpha
[2(c\! -\! 1)\! +\! 3q^2])}{4c_3{}^2\sqrt{2\! -\! c}},\nonumber\\
\end{eqnarray}
while $B_\theta=0$.
Equation (\ref{me1}) shows in a clear way that when the constant $c_4=0$ we have
no magnetic fields, namely the
magnetic fields are related to the integration constant   $c_4$.


\subsection{Rotating solutions}\label{S6}

 In this subsection  we   derive   rotating solutions  that satisfy
the field equations  (\ref{f1}), (\ref{f3}) and (\ref{maxf}).
In order to achieve this we apply the following
  transformation  \cite{Lemos:1994xp,Awad:2002cz}:
\begin{eqnarray}  \label{t1}
&&\bar{\phi} =\Xi~ {\phi}+\omega~t,\nonumber\\
&&
\bar{t}=
\Xi~ t+\omega~ \phi,
\end{eqnarray}
with  $\omega$  being the   rotation parameter and $\Xi=\sqrt{1+\omega^2}$.
Applying  (\ref{t1})
to the metric (\ref{met}) we obtain
\begin{eqnarray} \label{met1}
& &
\!\!\!\!\!\!\!\!\! \! \! \! \! \! \! \! \!
ds^2=[\Xi^2
H(r)-\omega^2r^2\sin^2\theta]dt^2-\frac{dr^2}{H(r)}
-r^2d\theta^2\nonumber\\
&&\!\!\!\!\!\!\!\!\! \! \!\! \!
- [ \Xi^2
r^2\sin^2\theta\!-\!\omega^2H]d\phi^2+2\omega
\Xi[H\!-\!r^2\sin^2\theta]dtd\phi,  \end{eqnarray}
   where $H(r)$ is given by the previously extracted static
solutions (\ref{sol11}), (\ref{sol22}), and $0\leq r< \infty$, $-\infty < t <
\infty$, $0 \leq \phi< 2\pi$. We mention that
the static configuration (\ref{met}) can be
recovered as a special case of the above general metric, if the rotation
parameter
$\omega$ is set to zero. Hence, for the general gauge potential
(\ref{pot1}) we acquire the form
\begin{equation}
\label{Rotpot}
\bar{V}=\left[\Xi q(r)+\omega s(r)\right]d\bar{t}+n(\phi) dr+\left[\omega q(r)+\Xi s(r)\right]d\bar{\phi}.
\end{equation}
Note here that although the transformation (\ref{t1}) leaves the local
properties of spacetime unaltered, it does change them globally as has been shown
in \cite{Lemos:1994xp}, since it mixes compact and noncompact coordinates. Thus,
the two metrics (\ref{met}) and (\ref{met1}) can be locally mapped into each
other but not globally \cite{Lemos:1994xp,Awad:2002cz}.

To conclude, we have succeeded to  derive  new rotating charged   black hole
solutions in
  $f({\cal R})$ gravity, using as a specific example the case
$f({\cal R})={\cal R}-2\alpha\sqrt{{\cal R}}$.
Similarly to the static case, these belong to two branches, one that contains
the   Kerr-Newman metric, namely the rotating charged black hole solution of
general
relativity, as a particular limit (the one arising inserting (\ref{sol11})
into  (\ref{met1})) and one that arises purely from the gravitational
modification and does not recover the general relativity solution (the one
arising inserting (\ref{sol22})
into  (\ref{met1})). Concerning  the singularity properties, as is clear
from   (\ref{met1}), these will be the same with the static solution
(\ref{met}).   Therefore,  at $r=0$ we obtain a
 true singularity, and close to $r=0$ the
behavior of the invariants are $(K,R_{\mu
\nu}R^{\mu \nu}) \sim {r^{-8}}$ and  $(R )\sim  {r^{-2 }}$.

\section{Thermodynamics   }\label{S66}

 In this section we focus on the investigation of the thermodynamic properties
  of the obtained  black hole solutions. Since solution (\ref{sol11}) contains
the general relativity result, in the following analysis we focus on the novel
solution  (\ref{sol22}) that arises solely from the gravitational modification \cite{Elizalde:2020icc,Nashed:2019yto,Nashed:2019tuk}.

We start by introducing the Hawking
temperature  as \cite{Sheykhi:2012zz,Sheykhi:2010zz,Hendi:2010gq,Sheykhi:2009pf}
  \begin{equation}
T_h = \frac{H'(r_h)}{4\pi},
\label{BHtemper}
\end{equation}
where the event horizon is located at $r = r_h$ which represents the largest positive root of $H(r_h) = 0$  that satisfies $H'(r_h)\neq 0$.
The Bekenstein-Hawking  entropy  of    $f({\cal R})$ gravitational theory is
given by \cite{Cognola:2011nj,Zheng:2018fyn}
\begin{equation}\label{ent}
S(r_h)=\frac{1}{4}Af_{R}(r_h),
\end{equation}
with  $A$  being the area of the event horizon. Additionally, the quasilocal
energy  in $f({\cal R})$ gravity is defined as
\cite{Cognola:2011nj,Zheng:2018fyn}
\begin{eqnarray}
\label{en}
&&
\!\!\!\!\!\!\!\!\!\!\!\!\!\!\!\!\!\!\!\!\!\!\!
E(r_h)=\frac{1}{4}  \int dr_h
\Big[r_h{}^2\Big\{f(R(r_h))-R(r_h)f_{R}(r_h)\Big\}\nonumber\\
&&
\ \ \ \ \ \ \ \ \ \ \ \
+
2f_{R}(r_h)
\Big].
\end{eqnarray}
 Finally, we can express the black hole mass as a function of the
horizon $r_h$    and the charge $q$, which
for the case  (\ref{sol22}),(\ref{me2}) becomes
\begin{eqnarray} \label{m33}
&&
{M_h}=\frac{r_h}{2}\left[\frac{c}{2}+\frac{q^2}{r_h{
}^2}\right].
\end{eqnarray}

\begin{figure}[ht]
\centering
  \includegraphics[scale=0.35]{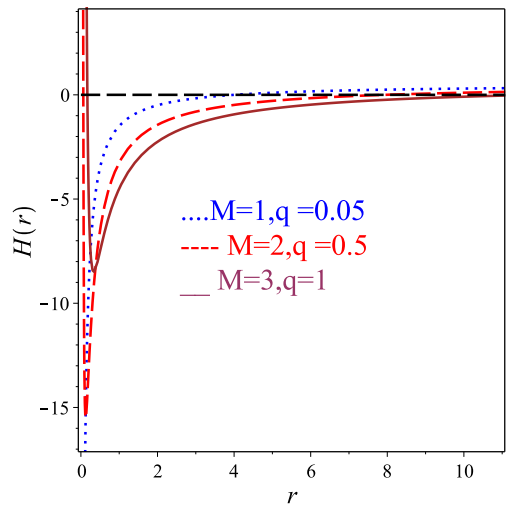}\hspace{0.2cm}
\caption{{\it{The metric function $H(r)$ vs the
radial coordinate  $r$,  for solution (\ref{sol22}) with $c=1$, for various
mass and charge
choices. The black hole  horizon is determined  by the condition
$H(r) = 0$. The ``degenerate horizon''   $r_{dg}$ marks the limiting value
after which there is no horizon and the central singularity becomes a naked
one (see text).}}}
\label{Fig:1}
\end{figure}

The  relation between the metric
function $H(r)$ and the radial coordinate $r$ is presented in  Fig.
\ref{Fig:1},
which shows the possible horizons of the solution. Note that  since  in 
this work for
simplicity we are using natural units, in order to be closer to physical cases
we should have taken   much larger values of $M$ and $q$ and then the radial
distance would take much larger values too while
$\alpha$ would take much smaller values. However, since in mathematical terms
the physical
properties of the solutions  do not depend on the scale,    and  in order to 
avoid graphs with very large/small numbers, we prefer to remain in these
representative numbers of order one since they are adequate in order to provide
the physical features of the solution. 

\begin{figure}[ht]
\centering
  \includegraphics[scale=0.35]{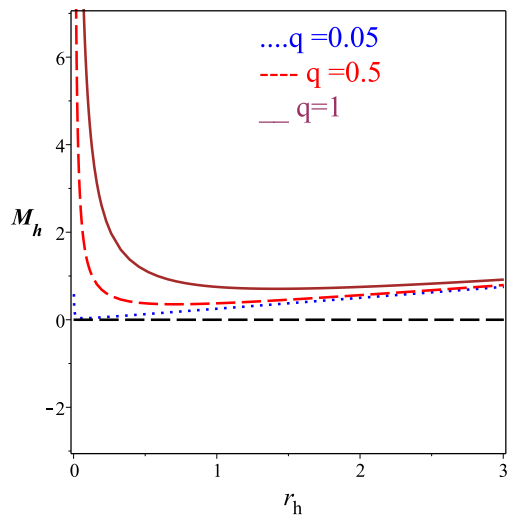}\hspace{0.2cm}
\caption{{\it{
The black hole mass as a function of the
horizon $r_h$, for solution (\ref{sol22}) with $c=1$, for various
  charge
choices. }}}
\label{Fig:2}
\end{figure}

Moreover, the relation
between $M_h$  and the horizon radius   is depicted
in Fig. \ref{Fig:2}. As one can see, there is a limiting horizon radius after
which there is no horizon and the black hole singularity will be a naked
singularities. This  ``degenerate horizon'' value $r_{dg}$ can thus be
calculated  by the condition   $\frac{\partial M_h}{\partial
r_h}$= 0,  which for solution  (\ref{sol22}),(\ref{me2}) yields $$r_{dg} =
\frac{\sqrt{2}q}{\sqrt{c}}.$$
Hence, as we can see, the cosmic censorship theorem can be violated in
non-linear Maxwell $f({\cal R})$ gravity.

\begin{figure}[ht]
\centering
  \includegraphics[scale=0.35]{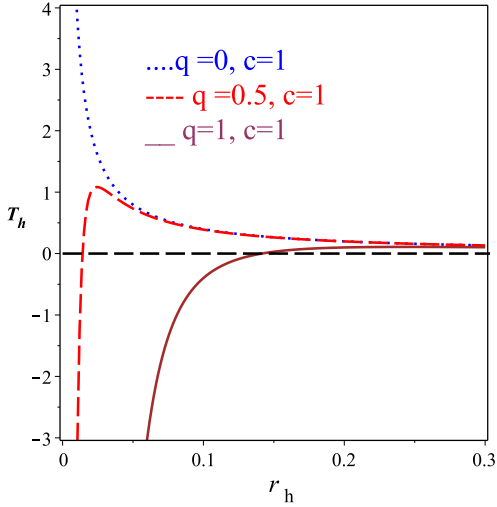}\hspace{0.2cm}
\caption{
{\it{The black hole temperature (\ref{m44temp}) as a function of the
horizon $r_h$, for solution (\ref{sol22}),  for solution (\ref{sol22}) with 
$c=1$, for various
  charge
choices.}}
}
\label{Fig:5}
\end{figure}

Concerning the Hawking temperature (\ref{BHtemper}), calculating it using
    the black hole solution
 (\ref{sol22}) we find
\begin{eqnarray} \label{m44temp}
{T_h}=\frac{cr_h{}^2-2q^2}{8\pi r_h{}^3}.
\end{eqnarray}
We mention that $T_h$ does not depend directly on the gravitational
modification parameter $\alpha$ (although it indirectly does since the latter
affects the horizon).
In  Fig. \ref{Fig:5} we depict the temperature behavior as a function of the
horizon.  As we observe, for suitable parameter values this can be negative,
and according to (\ref{m44temp}) this happens when
$c<\frac{2q^2}{r_h{}^2}$.
This implies a formation of an ultracold
black hole  \cite{Davies:1978mf,Babichev:2014lda}, which reveals the
capabilities of the scenario at hand.

\begin{figure}[ht]
\centering
  \includegraphics[scale=0.35]{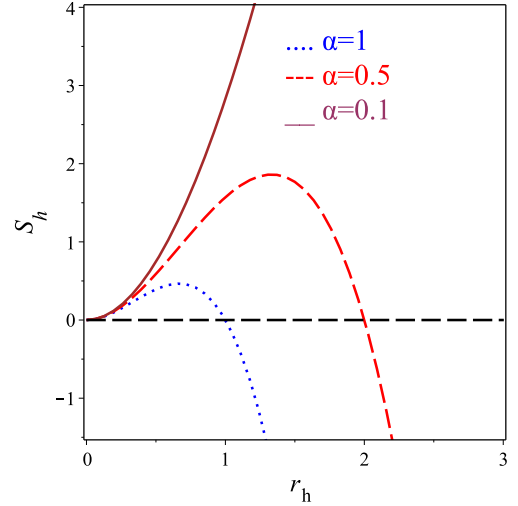}\hspace{0.2cm}
\caption{
{\it{The black hole entropy (\ref{ent1}) as a function of the
horizon $r_h$, for solution (\ref{sol22}),  for   various choices of the
gravitational modification parameter $\alpha$.}}
 }
\label{Fig:4}
\end{figure}

As a next step, using  expression (\ref{ent}) the entropy of the black hole
(\ref{sol22}) is  calculated  as
\begin{eqnarray}
\label{ent1}
{S_h}=\frac{\pi r_h{}^2(\sqrt{2-c}-\alpha r_h)}{\sqrt{2-c}}.
\end{eqnarray}
In Fig. \ref{Fig:4} we show the behavior of the entropy as a function of the
horizon. Hence, by imposing the  entropy positivity condition
we obtain
 $$\alpha<\frac{\sqrt{c-2}}{r_h}.$$
 This is one of the main results of the present work,
 and shows that the gravitational correction of $f({\cal R})$ gravity must be
 suitably small in order to avoid non-physical black hole properties (see
also the discussion in
\cite{Cvetic:2001bk,Nojiri:2001fa,Nojiri:2002qn,Nojiri:2017kex,Clunan:2004tb,
Nojiri:2001ae} for the entropy negativity in various theories of modified
gravity).

Similarly, using expression  (\ref{en}) we find the quasilocal energy of the
black hole  (\ref{sol22})  as
\begin{eqnarray}
\label{m44}
{E_h}=\frac{r_h(4\sqrt{2-c}+4r_h\alpha-3r_hc\alpha )}{8\sqrt{2-c}}.
\end{eqnarray}
From (\ref{m44}) we conclude that
in order to have  a positive value of the quasilocal energy we must have
\begin{equation}
\!\!\!\!\!\!\!\!\!\!\!\!   c>\frac{4}{3}\qquad  {\text{and}} \qquad
\alpha<\frac{4\sqrt{2-c}}{r_h(3c-4)}
\end{equation}
or
\begin{equation}
 \!\!\!\!\!\!\!\!\!\!\!\!   c<\frac{4}{3}\qquad   {\text{and}} \qquad
\alpha>\frac{4\sqrt{2-c}}{r_h(3c-4)}.
\end{equation}

\begin{figure}[ht]
\centering
  \includegraphics[scale=0.36]{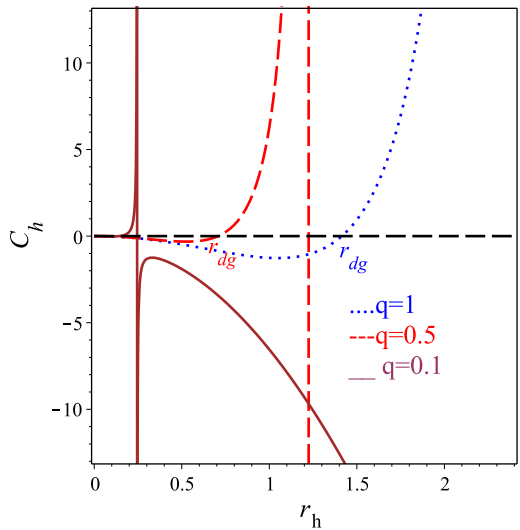}\hspace{0.2cm}
\caption{
{\it{The heat capacity (\ref{enr}) as a function of the
horizon $r_h$, for solution (\ref{sol22}) with $c=1$, for various
  charge
choices. }} }
\label{Fig:7}
\end{figure}

We continue by examining the  black hole
thermodynamical stability. As it is known, in order to analyze it one has to
examine the sign
of its heat capacity $C_h$, given as
\cite{Nouicer:2007pu,DK11,Chamblin:1999tk}:
\begin{equation}
\label{m55}
C_h=\frac{dE_h}{dT_h}= \frac{\partial M_h}{\partial r_h} \left(\frac{\partial
T_h}{\partial r_h}\right)^{-1},
\end{equation}
where $E_h$ is the energy.  If the heat capacity $C_{h} > 0$  then
the black hole is thermodynamically stable, i.e. a black hole
with a negative heat capacity is thermally unstable.
Concerning the heat capacity of the black hole solution (\ref{sol22}), using
Eq. (\ref{m55}) we acquire
\begin{equation}
\label{enr}
C_h=\frac{2\pi r_h{}^2(2q^2-cr_h{}^2)}{cr_h{}^2-6q^2}.
\end{equation}
We mention that $C_h$ does not depend directly on the gravitational
modification parameter $\alpha$, but only indirectly  through  the
effect of $\alpha$ on the horizon. This expression implies  that in order to obtain  a positive
heat capacity we must have
\begin{equation}\label{qcon}
q>\pm0.5r\,\sqrt{c}.\end{equation}
In Fig.
\ref{Fig:7} we depict $C_h$ as a function of the horizon, where we observe
that if $q$ satisfies the above inequality then stability is obtained.
We mention here that a negative heat capacity is
associated   with a negative temperature, which corresponds to $r_h <
r_{dg}$. At $r_h = r_{dg}$ both the temperature and the heat capacity are
exactly zero on the black hole horizon. When  $r_h > r_{dg}$, both temperature
and heat capacity are positive and the solution is in
thermal equilibrium. Indeed, the thermodynamical stability of   charged black
holes    has been widely studied
in various modified gravity theories, e.g. the thermodynamics of Bardeen
(regular) black holes \cite{Myung:2007qt}, of Schwarzschild-AdS solutions in
two vacuum scales case\cite{Dymnikova:2010zz}, of solutions in
noncommutative geometry
\cite{Berej:2006cc,Man:2013hpa,Tharanath:2014naa,Maluf:2018lyu}, etc.

\begin{figure}[ht]
\centering
  \includegraphics[scale=0.35]{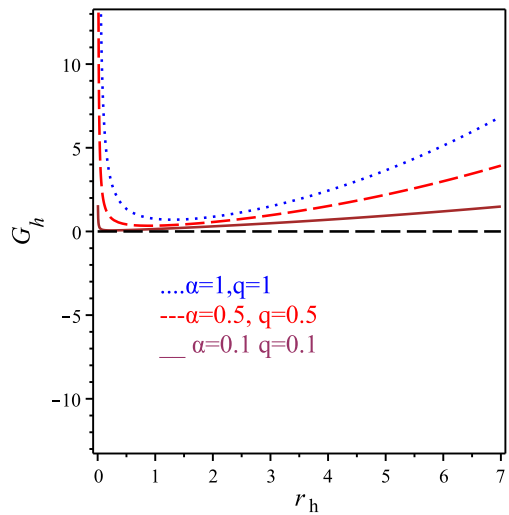}\hspace{0.2cm}
\caption{{\it{The black hole Gibb's free energy (\ref{m77}) as a function of
the
horizon $r_h$, for solution (\ref{sol22}) with $c=1$, for various choices of
charge and gravitational modification.  }}}
\label{Fig:8}
\end{figure}

Finally, let us make some comments on the Gibb's free energy, namely the free
energy in the grand canonical ensemble,
defined as \cite{Zheng:2018fyn,Kim:2012cma}
\begin{equation} \label{Gibbs}
G(r_h)=E(r_h)-T(r_h)S(r_h).
\end{equation}
Inserting   (\ref{m44temp}), (\ref{ent1}) and
(\ref{m44}) into (\ref{Gibbs}) we find
\begin{eqnarray} \label{m77}
&&{G_h}=\frac{(6q^2+cr_h^2)\sqrt{2-c}+\alpha r_h(r_h^2-2q^2)}{8r_h\sqrt{2-c}}.
\end{eqnarray}
The behavior of the Gibb's energy of the black holes (\ref{sol22}) is presented
in Fig.  \ref{Fig:8}  for particular values of the model parameters. As
we can see it is always positive when $\alpha>0$   which implies that it is more
globally stable.

\section{ Discussion and conclusion }
\label{S77}

The radical advance in multi-messenger astronomy opens the possibility to test
general relativity and investigate modified gravity by the gravitational and
electromagnetic waves    profile that arise  form mergers of spherically
symmetric objects, such as black holes and neutron stars. Hence, it is crucial
to study such object's properties in various theories of modified gravity in
the presence of the Maxwell sector.

In this work we investigated   static and rotating  charged
spherically symmetric  solution in the framework of    $f({\cal R})$
gravity, allowing additionally the electromagnetic sector to depart from
linearity. Applying a convenient, dual description for the electromagnetic
Lagrangian, and using as an example the square-root $f({\cal R})$   correction,
we were able to solve analytically the involved field equations. The
obtained solutions belong to two branches. One that contains
the   Kerr-Newman metric, namely the rotating charged black hole solution of
general relativity, as a particular limit  and one that arises purely from the
gravitational modification and does not recover the general relativity
solution.  Moreover, we have shown that the two components of the magnetic
fields, of the non-linear electrodynamics,  are connected by a constant which
if it is vanished we acquire a charged black hole with electric field only \cite{Nashed:2019tuk}.

Analyzing the novel black hole solution that does not have a general relativity
limit we found that it has a true  central singularity which is hidden behind
a horizon, however for particular parameter regions the horizon disappears and
the singularity becomes a naked one, i.e. we obtain a violation of the cosmic
censorship theorem.

Furthermore, we investigated the thermodynamical properties of the solutions,
such as the temperature, energy, entropy, heat capacity and Gibbs free energy.
We extracted the conditions on the gravitational modification parameter in
order to obtain entropy and quasilocal energy positivity. Concerning
temperature,
we showed that it can become negative for particular parameter
values, and thus   ultracold black holes may be formed. Finally, we examined
the thermodynamic stability of the solutions by examining the sign of the heat
capacity, extracting the corresponding conditions.

In summary, we showed that even small deviations from general relativity
and/or from linear electrodynamics may lead to novel spherically symmetric
solution branches, with novel properties that do not appear in standard
general relativity. Since these properties may be embedded in the gravitational
waves profiles, they could serve as a smoking gun of this subclass of
gravitational modification.

\section*{Acknowledgements}
  This article is based upon work from CANTATA COST (European Cooperation in
Science and Technology) action CA15117, EU Framework Programme Horizon 2020.
It is supported in part by the USTC Fellowship for international
professors.


\end{document}